\documentstyle[eqsecnum,preprint,prd,aps]{revtex}

\begin{document}\bibliographystyle{prsty}
\draft
\preprint{ADP-93-199/M14}
\title{
The quantum behavior of general time dependent\\ quadratic systems
linearly coupled to a bath}
\author{J. Twamley}
\address{
Department of Physics and Mathematical Physics, University of Adelaide\\
GPO Box 498, Adelaide, South Australia 5001}
\date{\today}
\maketitle
\begin{abstract}
In this paper we solve for the quantum propagator of a
general time dependent system quadratic in both position and
momentum, linearly coupled to an infinite bath of harmonic
oscillators. We work in the regime where the quantum optical
master equation is valid. We map this master equation to a
Schroedinger equation on Super-Hilbert space and utilize Lie
Algebraic techniques to solve for the dynamics in this space. We
then map back to the original Hilbert space to obtain the solution of
the quantum dynamics. The Lie Algebraic techniques used are
preferable to the standard Wei-Norman methods in that only
coupled systems of first order ordinary differential equations and
purely algebraic equations need only be solved. We look at two
examples.
\end{abstract}
\narrowtext

\section{Introduction}
In this note we describe the solution
of the quantum optical
master equation for a general time dependent system at most
quadratic in position and momentum. The coupling to the bath is
taken to be a linear  and the bath is comprised of
harmonic oscillators.  In the optical regime we can write an
approximate quantum master equation \cite{GARDINER:QUANTUM_NOISE}
\begin{equation}
\dot{\rho}=-\frac{i}{\hbar}[H,\rho]+\Lambda\rho\;\;,\label{eq1}\end{equation}
\begin{equation}
{\bf\Lambda}\rho=
\frac{\gamma(\bar{n}+1)}{2}\left\{[a ,\rho a^{\dagger} ]+[a\rho,a^{\dagger} ]
\right\}
+\frac{\gamma\bar{n}}{2}\left\{[a^{\dagger} ,\rho a ]+[a^{\dagger}\rho
,a ]\right\}\;\;,\label{eq2}\end{equation}
where $\gamma$ is the strength of the coupling to the bath and $\bar{n}=[\exp
(-\beta\hbar\omega)-1]^{-1}$ is the equilibrium number of
photons in the bath. The
approximations necessary to obtain equation (\ref{eq1}) are: rotating
wave approximation (RWA), small system anharmonicity,
Markovian evolution, weak coupling and initially no correlations
between the system and  bath.
It is generally accepted that equation
(\ref{eq1}) describes well the physics near optical frequencies and
forms the basis of many models studied in quantum optics
\cite{GARDINER:QUANTUM_NOISE,LOUISELL:QUANTUM_STATISTICAL}
(and references therein). We take as
our system the Hamiltonian \begin{equation} H/\hbar=\omega (t)\,a^{\dagger}
a+f_{1}(t) \,a+\bar{f}_{1}(t)\,a^{\dagger}+
f_{2}(t)\,a^{2}+\bar{f}_{2}(t)\,a^{\dagger}{}^{2}\;\;,
\label{eq3}\end{equation}
where
$\omega, f_{1},f_{2},\bar{f}_{1},\bar{f}_{2}$ are arbitrary functions
of time.  We will assume $\omega\in{\rm\bf R}$ but will not
necessarily assume that $f_{1}^{*}=\bar{f}_{1}$ and
$f_{2}^{*}=\bar{f}_{2}$. This is to accommodate the phenomenological
modeling of dissipation through a non-unitary Hamiltonian
\cite{DATTOLI_TORRE:1990}. The general form of the system (\ref{eq3})
covers many specific models including the simple harmonic
oscillator, the parametric oscillator, the forced harmonic oscillator
and the degenerate parametric oscillator. The isolated system also
exhibits many families of coherent states ({\it a la} Perelomov
\cite{PERELOMOV:GENERALISED_COHERENT}) for specific functional forms of
$\omega,f_{i}$ and
$\bar{f}_{i}$. It is of interest to study the effects of the bath on
these coherent states. In particular, in the absence of the bath, the
coherent state evolves solitonically under the action of a
Hamiltonian which is a linear combination of the generators of the
Lie group defining the coherent state \cite{GERRY:1985}. The coherent
state is very  special  in that it follows the exact classical equations
of motion forever.  However,  since coherent states still display
quantum interference  it is incorrect to consider such states as
truly classical. Much work has been done recently on the {\em
decoherence} effects of an external bath and the suppression of
quantum interference in the system \cite{ZUREK:1991}.
More important is the study of the effects of the bath on quantum
systems which, when isolated from the bath, do {\em not} possess
coherent states. Does the interaction with the bath succeed in
retarding the spread of an initial wave packet? ie. can one
interpret the bath as effectively measuring the system and
``collapsing'' the wave function of the system? Does the interaction
with the bath cause a deviation from the quantal equations of
motion towards the classical equations of motion? As a first step
towards answering these important questions one must solve for
the quantum behavior of the system coupled to the bath. We do
so in the quantum  optical limit for a general system linearly
coupled to the bath.

The standard method used to solve equation (\ref{eq1}) is to
transform the master equation into a c--numbered partial
differential equation for a particular quantum distribution
function (QDF). However, in certain important cases (degenerate
parametric oscillator in the P representation), the resulting PDE
can take the form of a Fokker-Planck equation with a negative
diffusion matrix. Such cases do not correspond to the usual
diffusive process. To use the machinery associated with diffusive
processes one must make use of the generalized
P--Representation \cite{DRUMMOND_GARDINER:1980}. This entails doubling the
degrees of freedom and ``choosing a gauge'' to ensure that the
diffusion matrix is positive and real. However, recent numerical
work using this QDF has raised questions concerning  it's validity
\cite{SMITH_GARDINER:1989}.  Of course, for system non-linearities higher than
quadratic one has a c--numbered PDE  of order three or
more. Since this does not occur in this model we will not address
this here. Resolving (\ref{eq2}) onto the $x,p$ basis via $a=(\omega
x+ip)/\sqrt{2\hbar\omega}$ we get
\begin{eqnarray} 4\hbar\omega{\bf\Lambda}\rho/\gamma &=&
-\omega^{2}(2\bar{n}+1)[x,[x,\rho]]-(2\bar{n}+1)[p,[p,\rho]]\nonumber\\
& &+i\omega[p,\rho x+x\rho]-i\omega[x,\rho p+p\rho]\;\;,\label{master1}
\end{eqnarray}
while resolving (\ref{eq1}) onto the Glauber-Cahill family of QDFs
$W(\alpha,\alpha^{*},s)$, which include the P (s=+1), Wigner (s=0)
and Q (s=-1) quantum distribution functions we obtain
\widetext
\begin{eqnarray}
\dot{W}(\alpha,\alpha^{*},s) = &&
  -\frac{i}{\hbar}\left\{[f_{1}+2\alpha f_{2}+\alpha^{*}\omega]
\partial_{\alpha^{*}}-[\bar{f}_{1}+2\alpha^{*}\bar{f}_{2}+\alpha\omega]
\partial_{\alpha}
 -s[f_{2}-\bar{f}_{2}]\partial^{2}_{\alpha\alpha^{*}}\right\}
W(\alpha,\alpha^{*},s)\nonumber\\
&&+\frac{\gamma}{2}\left\{\partial_{\alpha}\,\alpha +\partial_{\alpha^{*}}
\,\alpha^{*}+(2\bar{n}+1-s)\partial^{2}_{\alpha\alpha^{*}}\right\}
W(\alpha,\alpha^{*},s)\;\;.\label{Cahill-glauber}\end{eqnarray}
{}From (\ref{master1}) we see
that we now are effectively coupled to noise in both position and
momentum.  We now convert equation (\ref{eq1}) into a
Schroedinger equation on Super--Hilbert space.

\narrowtext
\section{Super--Hilbert Space and Lie Algebraic Solutions}
We initially follow a technique developed
by Barnett and Knight
in association with the applications of Thermofield theory to
quantum optics \cite{BARNETT_KNIGHT:1985}.  In this theory, the
density operator is transformed into a state vector in an
expanded Hilbert space (Super-Hilbert space) and operations on
operators become super-operators in the Super-Hilbert space.
The transformation between Hilbert space and Super-Hilbert
space is accomplished through the state
\begin{equation} |I\rangle\equiv \sum_{N} |N,N\rangle\equiv\sum_{N}|N
\rangle_{1}\times
|N\rangle_{2}\;\;,\label{I-DEF}\end{equation}
and through defining $|\rho\rangle=\rho\,|I\rangle$.
One can show $a|I\rangle=\tilde{a}^{\dagger}|I\rangle$ and
$a^{\dagger}|I\rangle=\tilde{a}|I\rangle$ where the $a$'s are the annihilation
operators in Hilbert space 1 (the original space) and the
$\tilde{a}$'s are the annihilation operators in Hilbert space 2.
Using (\ref{I-DEF}), equation (\ref{eq1}) becomes a Schroedinger like
equation on the Super-Hilbert space
\begin{equation} \frac{d}{dt}|\rho\rangle=-i\tilde{H}|\rho\rangle\;\;,
\label{schro2}\end{equation}
where
\widetext
\begin{eqnarray} -i\tilde{H} && =
-i\left\{
\omega\,[a^{\dagger} a -\tilde{a}^{\dagger} \tilde{a} ]+f_{1}\,[a
-\tilde{a}^{\dagger} ]+\bar{f}_{1}\,[a^{\dagger} -\tilde{a} ]+f_{2}[a^{2}
-\tilde{a}^{\dagger\, 2} ] +\bar{f}_{2}[a^{\dagger\, 2}
-\tilde{a}^{2} ]\right\}\nonumber\\
&& +\frac{\gamma}{2}(\bar{n}+1)[2a \tilde{a} -a^{\dagger} a
-\tilde{a}^{\dagger}
\tilde{a} ]
+\frac{\gamma}{2}\bar{n}[2a^{\dagger} \tilde{a}^{\dagger} -a a^{\dagger}
-\tilde{a} \tilde{a}^{\dagger} ]\;\;.\label{schro}\end{eqnarray}
\narrowtext
The ``Super-Hamiltonian'', $\tilde{H}$ seems at first glance
 totally intractable, however, upon inspection one can rewrite
(\ref{schro2}) as
\begin{equation} \frac{d}{dt}U(t)=-iH(t)\,U(t)\qquad\qquad
|\rho(t)\rangle\equiv
U(t)|\rho(0)\rangle\;\;,\label{schro3}\end{equation}
\begin{equation} H(t)=\sum_{i=0}^{N}\;h_{i}(t)\,N_{i}\;\;,\label{schro4}
\end{equation}
where the $N_{i}$ are the generators of an $N$ dimensional Lie
algebra $\cal L$. We therefore have access to powerful Lie Algebraic
techniques to solve (\ref{schro3}) for the propagator $U(t)$.

The algebra $\cal L$ can be identified as the two photon subgroup of
Sp(6,R). This subgroup is a 15
dimensional, semisimple Lie algebra, possessing a five dimensional
ideal $\cal I$,  $\{N_{0},N_{1},N_{2},N_{3},N_{4}\}$ and compact $U(2)$
subgroup $\{N_{8},N_{9},N_{13},N_{14}\}$. It also contains the
single photon algebra $h_{4}$ and Weyl--Heisenberg algebra
$h_{3}$ for both sets of oscillators. This algebra has been studied in
detail by Gilmore and Yuan \cite{GILMORE_YUAN:1989}.
The description of the generators
$N_{i}$ and coefficient functions $h_{i}(t)$ are given in Table \ref{table1}
and Table \ref{table2}.

The standard Lie Algebraic method used to solve equations such as
(\ref{schro3}) is the Wei--Norman method \cite{WEI_NORMAN:1963}. In
this method one assumes a particular form for the propagator as
an ordered product of exponentials of the generators
\begin{equation} U(t)=\prod_{i=0}^{N}\,\exp (g_{i}(t)\,N_{i})\;\;.
\label{ansatz}\end{equation}
 One
then substitutes this ansatz into (\ref{schro3}) to obtain differential
equations relating $g_{i}$ to $h_{i}$. However, the resulting
equations are highly non-linear and coupled. The complexity of
these equations and the effort needed to obtain them grows
rapidly with the dimension of the Lie group and proves
prohibitive for semisimple Lie groups with dimensions greater
than six. Even if one succeeds in solving the  coupled ODEs for the
$g_{i}$'s the global validity of the chosen ansatz is not guaranteed
and must be tested through examination of the $g_{i}$'s
\cite{WEI_NORMAN:1963b}. Finally, the whole procedure is highly sensitive
to the particular ordering of generators chosen in the ansatz for
$U(t)$.  We will, instead, adopt a method discovered by Fernandez
\cite{FERNANDEZ:1989}, which can reduce the problem to the solution of
a coupled set of {\em linear} ODEs and algebraic equations. Key to
this approach is the existence of a suitably large proper ideal $\cal I$. One
effectively solves for the dynamics of the ideal and ``lifts'' this
information via Baker--Campbell--Hausdorff (BCH)
 identities to determine the  dynamics of the
entire algebra.

For any element of the proper ideal $N_{i}\in\cal I$ ($i\in \{0,..,4\}$)
we have $U^{-1}{\cal I}U\in\cal I$ where $U$ is the propagator in
(\ref{schro3}). In particular
\begin{equation} N_{i}(t)\equiv
U^{-1}(t)N_{i}U(t)=\sum_{j=0}^{4}\,u_{ij}(t)\,N_{j}\;\;,\qquad
i\in\{0,..,4\}\;\;.\label{u-def}\end{equation}
It is easy to show that the time dependent coefficients $u_{ij}(t)$
obey
\begin{equation} \dot{u}=i{\cal H}{\bf\cdot}u\;\;,\;\;
{\cal H}_{ij}(t)=\sum_{m=1}^{N}\,h_{m}(t)\,C_{mj}^{k}\;\;,\;\;
u_{ij}(t=0)=\delta_{ij}\;\;,
\label{u-def2}\end{equation}
where $C_{ij}^{k}$ are the structure functions of $\cal L$.
To solve for $u_{ij}(t)$ we must solve the linear set of coupled
ordinary differential equations (\ref{u-def2}).  By transforming to a
new set of generators for $\cal L$ we can simplify the structure of
(\ref{u-def2}) greatly. We will denote this new set by $\{W_{i}\}$. This
change of basis decouples (\ref{u-def2}) into pairs of coupled linear
first order differential equations and also greatly simplifies the
BCH disentangling for the relations between the $u_{ij}$'s and the
coefficient functions $g_{i}(t)$ appearing in (\ref{ansatz}).
To obtain the time dependent coefficients $g_{i}(t)$ appearing in
(\ref{ansatz}) we use BCH disentangling identities
to compute the functionals $F_{j}(g_{i})$ appearing in
\begin{equation}
U^{-1}(t)N_{i}U(t)=\sum_{j=0}^{4}\,F_{j}(g_{i})N_{j}\;\;.
\label{resolve}\end{equation}
The $F_{j}$'s are algebraic functions of the $g_{i}$'s. Equating
the coefficients of $N_{j}$ in (\ref{u-def}) and (\ref{resolve})
for a particular
$N_{i}$ yields algebraic relations between the $g_{i}$'s and the $u_{ij}$'s.
To check the global validity of the particular ansatz chosen one must
only determine when the algebraic relations become degenerate.

Before we solve for the dynamics in the extended Super--Hilbert space
we show first how to recover the dynamics of the density operator in the
original Hilbert space. Resolving the density operator into number
states gives
\begin{equation}
\rho=\sum_{nm}\,\rho_{nm}\,|n\rangle\langle m|\;\;,\qquad\rho_{nm}=\langle
n|\rho |m\rangle\;\;.\label{resolve1}\end{equation}
We obtain the density {\em state} vector in the extended Hilbert space
through
\begin{equation}
|\rho\rangle\equiv\rho|I\rangle =\sum_{nm}\,\rho_{nm}|n\rangle\times |m\rangle
= \sum_{nm}\,\rho_{nm}|n,m\rangle\;\;.\end{equation}
Thus, the original density operator may be recovered through
\begin{equation} \rho=\sum_{nm}\,\langle m,n|\rho\rangle |n\rangle\langle
m|\;\;.\label{recover}\end{equation}
Through similar steps one can also recover the components of the
original density operator in the coherent state basis. We resolve
unity in the coherent state basis through
$I=\frac{1}{\pi}\int\,d^{2}\alpha\,|\alpha\rangle\langle\alpha |$ and using
the identity
\begin{equation} \sum_{n=0}^{\infty}\,\langle\beta |n\rangle\,|n\rangle =
|\beta^{*}\rangle\;\;,\end{equation}
we obtain
\begin{equation} |\rho\rangle =\frac{1}{\pi^{2}}\int d^{2}\alpha\,d^{2}\beta\,
\rho\,(\alpha^{*},\beta)\,e^{-(|\alpha|+|\beta|)/2}\,|\alpha,\beta^{*}
\rangle\;\;,\label{coherent-resolve}\end{equation}
where we have set
$|\alpha,\beta\rangle\equiv|\alpha\rangle\times |\beta\rangle$ and
$\rho\,(\alpha^{*},\beta)=\langle\alpha|\rho|\beta\rangle$. One can
easily show
\begin{equation}
\langle\alpha|\rho|\beta\rangle =\langle\beta^{*},\alpha|\rho\rangle\;\;,
\label{coherent-resolve1}\end{equation}
as expected. From (\ref{coherent-resolve1}) we can obtain the Q
distribution function \cite{HILLERY_OCONNELL:1984}
\begin{equation}
Q(\alpha^{*},\alpha,t)=\langle\alpha|\rho(t)|\alpha\rangle =
\langle\alpha^{*},\alpha
|\rho(t)\rangle\;\;.\label{coherent-resolve2}\end{equation}
However, in the extended Hilbert space,
$|\rho(t)\rangle =U(t)|\rho(0)\rangle$. Using this with equations
(\ref{coherent-resolve}) and  (\ref{coherent-resolve2}) we finally obtain
\begin{eqnarray}
Q&&(\alpha^{*},\alpha,t)= \nonumber\\
&&\frac{1}{\pi^{2}}\int d^{2}\alpha\,d^{2}\beta\,
\rho_{0}(\tilde{\alpha}^{*},\tilde{\beta})\,e^{-(|\alpha|+|\beta|)/2}
\langle\tilde{\alpha}^{*},\tilde{\alpha}|U(t)|\alpha,\beta^{*}\rangle
\;\;.\label{coherent-resolve4}\end{eqnarray}
In the examples treated in this paper the coherent state basis proves
more useful than the number basis. Thus to calculate the Q function it
will be necessary to compute
\begin{equation}
\langle\tilde{\alpha}^{*},\tilde{\alpha}|U(t)|\alpha,\beta^{*}\rangle
\;\;,\label{coherent-resolve5}\end{equation}
for various propagators $U(t)$. In the systems treated, $U$ will be at
most quadratic in the annihilation and creation operators of the two
Hilbert spaces. Writing
\begin{equation} |\alpha^{*},\beta\rangle =e^{-|\alpha|/2-|\beta|/2}\,
e^{\alpha\,a^{\dagger}+\beta^{*}\,\tilde{a}^{\dagger}}\,|0,0\rangle\;\;,
\end{equation}
equation (\ref{coherent-resolve5}) becomes $\langle
0,0|\tilde{U}(t)|0,0\rangle$ where
\begin{equation} \tilde{U}(t)=e^{-|\tilde{\alpha}|-(|\alpha|+|\beta|)/2}\;
e^{\tilde{\alpha}^{*}a+\tilde{\alpha}\tilde{a}}\,U(t)\,
e^{\alpha
a^{\dagger}+\beta^{*}\tilde{a}^{\dagger}}\;\;.\label{coherent-resolve6}
\end{equation}
To evaluate the above we use BCH identities to normal order the
operator $\tilde{U}$ as
\begin{equation} \tilde{U}(t)=e^{ R_{ij}a^{\dagger}_{i}a^{\dagger}_{j}
+c_{i}a^{\dagger}_{i}}\,
e^{D_{ij}(a^{\dagger}_{i}a_{j}+\delta_{ij}/2)+\eta I}\,
e^{L_{ij}a_{i}a_{j}+l_{i}a_{i}}\;\;,\label{coherent-resolve7}\end{equation}
where the index i labels the two Hilbert spaces.
The vacuum expectation value of $\tilde{U}$ becomes trivial and yields
\begin{equation} \langle 0,0|\tilde{U}(t)|0,0\rangle=\exp
(\frac{D_{11}+D_{22}}{2}+\eta)\;\;.\end{equation}

\section{Solving the Dynamics}
In Table \ref{table3}
we show the relation between $N_{i}$ and $W_{i}$ and in Table \ref{table4} we
give the commutation table for the ideal $\cal I$.

{}From the $\{W_{i}\}$ commutation table we can construct the
``Hamiltonian'' $\cal H$ in (\ref{u-def2})
\mediumtext
\begin{equation}{\cal H}_{ij}=\left[
\begin{array}{ccccc}
0 ,& 0, & 0, & 0, & 0, \\
-2h_{4}, & -h_{8}/4-h_{9}/2, & -2h_{6}, & 0, & 0 \\
-2h_{3} ,& -2h_{5}, & -h_{8}/4+h_{9}/2, & 0, & 0, \\
2h_{2} , & 2h_{7}, & 0 ,& h_{8}/4 -h_{9}/2 ,& 2h_{6}, \\
2h_{1}, & 0, & 2h_{7}, & 2h_{5}, & h_{8}/4+h_{9}/2, \\
\end{array}\right]
\end{equation}
\narrowtext
We see that the degree of coupling within (\ref{u-def2}) is related to the
number of nonzero entries in each row of ${\cal H}_{ij}$. This is in
turn related to the number of different $W_{i}$'s appearing in each
row of the commutation table of the ideal $\cal I$.
{}From the initial condition $u_{ij}(t=0)=\delta_{ij}$  and
the structure of ${\cal H}_{ij}$ we immediately obtain
\begin{equation} u_{i0}=\delta_{i0}\;\;,\qquad
    u_{13}=u_{23}=u_{14}=u_{24}=0\;\;\forall
t\;\;.\label{initial-zeros}\end{equation}
Thus the set (\ref{u-def2}) of ODE's for $u_{1i}$ and
$u_{2i}$ can be re-cast as
\begin{eqnarray} \dot{X}_{1} & = & i{\bf A}X_{1}+i{\bf B}\;\;,
\label{big-equ1}\\
\dot{X}_{2} & = & i{\bf C}X_{2}+i{\bf D}\;\;,\label{big-equ2}\end{eqnarray}
where
\begin{eqnarray} X_{1}&\equiv&\left(\begin{array}{ccc}
u_{10} & u_{11} & u_{12} \\
u_{20} & u_{21} & u_{22} \end{array} \right)\;,\\ \noalign{\medskip}
 X_{2}&\equiv&\left(\begin{array}{ccccc}
u_{33} & u_{34} & u_{31} & u_{32} & u_{30} \\
u_{43} & u_{44} & u_{41} & u_{42} & u_{40} \end{array}\right)\;\;,
\end{eqnarray}

\begin{eqnarray}
{\bf A}&\equiv&\left(\begin{array}{cc} {\cal H}_{11} & {\cal H}_{12} \\
{\cal H}_{21} & {\cal H}_{22} \end{array}\right) =
\left(\begin{array}{cc}
-i\gamma/2-\omega\;, & 2\bar{f}_{2} \\
-2f_{2}\;, & -i\gamma/2+\omega \end{array} \right)\;,\\ \noalign{\medskip}
{\bf C}&\equiv&\left(\begin{array}{cc}
{\cal H}_{33} & {\cal H}_{34} \\
{\cal H}_{21} & {\cal H}_{44} \end{array}\right)=\left(\begin{array}{cc}
-i\gamma/2-\omega & 2\bar{f}_{2} \\
-2f_{2} & i\gamma/2+\omega \end{array} \right)\;,\\ \noalign{\medskip}
{\bf B}&\equiv&\left(\begin{array}{ccc}
{\cal H}_{10} & 0 & 0 \\
{\cal H}_{20} & 0 & 0 \end{array}\right)\;\;,\\ \noalign{\medskip}
{\bf D}&\equiv&\left(\begin{array}{ccccc}
0 & 0 & {\cal H}_{31}u_{11} & {\cal H}_{31}u_{12} & {\cal H}_{30} \\
0 & 0 & {\cal H}_{42}u_{21} & {\cal H}_{42}u_{22} & {\cal H}_{40}
\end{array}\right)\;\;.
\end{eqnarray}

{}From (\ref{u-def2}), the initial value of $X_{1}$ and $X_{2}$ are
\begin{eqnarray} X_{1}(t=0)&=&\left(\begin{array}{ccc}
0 & 1 & 0 \\
0 & 0 & 1 \end{array} \right)\;,\\ \noalign{\medskip}
X_{2}(t=0)&=&\left(\begin{array}{ccccc}
1 & 0 & 0 & 0 & 0 \\
0 & 1 & 0 & 0 & 0 \end{array} \right)\;\;.\end{eqnarray}
However, from Table \ref{table3}, $h_{4}=h_{3}=0$ and thus $\bf B$ vanishes
identically. Equation (\ref{big-equ1}) then gives
\begin{equation} u_{10}=u_{20}=0\;\;\forall t\label{X1-solution}
\;\;.\end{equation}
Note, however, that equation  (\ref{big-equ1}) must already
be solved to correctly specify $\bf D$ in (\ref{big-equ2}).
The solution of equations
(\ref{big-equ1},\ref{big-equ2})
together with (\ref{initial-zeros}) completely determine the dynamical
evolution of the $u_{ij}(t)$.

To finally compute  the propagator we must
calculate the algebraic relations between the $g_{i}(t)$ and
$u_{ij}(t)$. We must evaluate the action of the group on each of the
elements of the ideal, that is, we must calculate
\begin{equation} \exp (-g_{i}\,{\rm ad}\, W_{i})\, W_{j}\equiv
e^{-g_{i}W_{i}}\,
W_{j}\, e^{g_{i}W_{i}}\;\;,\end{equation}
for $i\in (0,..,14)\;,\;\; j\in (0,..,4)$.
This is achieved using Baker--Campbell-Hausdorff disentangling
identities for Lie Groups. The relevant identities are given in
Appendix 1. Tables \ref{table5} and \ref{table6} summarize the
action of the group on the
elements of the ideal.
The particular set of algebraic relations obtained will depend
sensitively on the choice of ordering of the generators in the ansatz
for $U(t)$. We choose the ordering
\begin{equation} U(t)=\tilde{\prod}_{i}\,\exp ( g_{i}(t)W_{i})\;\;,
\label{prop-ansatz}\end{equation}
where $\tilde{\prod_{i}}$ is the product ordered from the left in the
sequence
1, 2, 13, 5, 3, 11, 7, 8, 9, 14, 10, 6, 12, 4, 0.
With this ordering the six conditions
(\ref{initial-zeros}, \ref{X1-solution}) give
\begin{equation} g_{4}=g_{3}=g_{14}=0\;\;,\qquad g_{10}=-2g_{12}\;,\qquad
g_{11}=-2g_{13}\;\;.\end{equation}
The very complicated disentangling of $W_{14}$ is thus avoided as
$g_{14}=0$ with this choice of ordering. Working through the BCH identities
with a symbolic manipulator
we finally arrive at the relations in Table \ref{table7}.

The relations are analytic except when $u_{11}=0$ or $u_{33}=0$. Thus
the ansatz (\ref{prop-ansatz}) is global if both $u_{11}\neq 0$ and
$u_{33}\neq 0$ for all $t$. Except for $g_{0}(t)$, the propagator is
formally determined. To solve for $g_{0}(t)$ however, is not trivial.
It cannot be determined through BCH identities as $W_{0}$ commutes
with every element of the group. Instead we must substitute
(\ref{prop-ansatz}) into the ``Schroedinger'' equation ${\cal H}
=i\dot{U}U^{-1}$ and from this determine $g_{0}$. In
\cite{FERNANDEZ:1989} the resulting relation (just below equation (20) in
\cite{FERNANDEZ:1989}) is quite simple. However, for the much larger algebra
Sp(6,r) the resulting relation can be quite complicated. To compute
the above ``Schroedinger'' equation and obtain the general relation
for $g_{0}$ it
is expedient to use the faithful matrix representation of Sp(6,r)
given by Gilmore and Yuan \cite{GILMORE_YUAN:1989}. We will
not give the general formula but will compute it case by case.

\section{Examples}
In this section we apply the above methods to solve two
model systems. The first is the simple harmonic oscillator. We set
$\omega(t)\equiv\omega$ and $f_{1}=f_{2}=\bar{f}_{1}=\bar{f}_{2}=0$.
We compute the matrix ${\cal H}_{ij}$ in (\ref{u-def2}) and construct the
matrices $\bf A,B$ and $\bf C$. We can easily solve the coupled sets
of ODEs for $X_{1}$ and $X_{2}$ to find the non-vanishing $u_{ij}$ to be
\begin{equation}
u_{11}=u_{22}^{*}=u_{33}^{-1\,*}=u_{44}^{-1}=e^{(-\gamma/2+i\omega)t}
\;\;,\end{equation}
\begin{equation} u_{31}=-2(2\bar{n}+1)e^{-i\omega t}\sinh \frac{\gamma
t}{2}\;,\qquad\qquad
u_{42}=e^{2i\omega t}u_{31}\;\;.\end{equation}
The resulting non-zero $g$'s are
\begin{equation} g_{7}=(2\bar{n}+1)e^{-\gamma/2 t}\sinh \frac{\gamma
t}{2}\;,\;\;
g_{8}=2\gamma t\;,\;\;
g_{9}=-2i\omega t\;\;.\end{equation}
The propagator on the extended Hilbert space is
$U(t)=e^{g_{7}W_{7}}\,e^{g_{8}W_{8}}\,e^{g_{9}W_{9}}e^{g_{0}W_{0}}$.
This may be
normal ordered in terms of SU(1,1) generators \cite{WODKIEWICZ_EBERLY:1985},
\[ K_{+}=a^{\dagger}\tilde{a}^{\dagger}\;\;,\]
\[ K_{-}=a\tilde{a}\;\;,\]
\[ K_{3}=(a^{\dagger}a+\tilde{a}\tilde{a}^{\dagger})/2\;\;,\]
\[ K_{0}=a^{\dagger}a-\tilde{a}^{\dagger}\tilde{a}\;\;,\]
to give
\[ U(t)=\exp (xK_{+})\exp (yK_{3}+g_{9}/2K_{0})\exp (xK_{-})\exp(g_{0}W_{0})\]
 where
\begin{eqnarray} x & = &
\frac{g_{7}e^{g_{8}/4}-\sinh\,g_{8}/4}{g_{7}e^{g_{8}/4}+\cosh\,g_{8}/4}\;\;,\\
y & = & -2\ln\,\left[g_{7}e^{g_{8}/4}+\cosh\,g_{8}/4\right]\;\;,\\
z & = &
\frac{g_{7}e^{g_{8}/4}+\sinh\,g_{8}/4}{g_{7}e^{g_{8}/4}+\cosh\,g_{8}/4}\;\;.
\end{eqnarray}
To obtain $g_{0}$ we compute ${\cal H}=i\dot{U}U^{-1}$ to find
$\dot{g_{0}}=\gamma/2$.
With $U$ normal ordered and the identity \cite{LOUISELL:QUANTUM_STATISTICAL},

\[ [\exp(-xa^{\dagger}a)]_{\rm normal}=\sum_{l=0}^{\infty}
\frac{(e^{-x}-1)^{l}}{l!}\,a^{\dagger\,l} a^{l}\;\;,\]   it
is a simple matter to compute (\ref{coherent-resolve5}),
\begin{eqnarray}\langle\tilde{\beta}^{*},\tilde{\alpha}|&&U(t)|
\alpha,\beta^{*}\rangle= \nonumber\\
 &&\frac{1}{\kappa}\langle\tilde{\alpha}|\tilde{\beta}\rangle
\langle\beta |\alpha\rangle
\exp\left[-\frac{1}{\kappa}
(\tilde{\alpha}-\beta\chi)^{*}(\tilde{\beta}-\alpha\chi)\right]\;\;,
\label{prop-alpha}\end{eqnarray}
where $\kappa\equiv 1+\bar{n}(1-e^{-\gamma t})$ and $\chi\equiv
\exp(-(\gamma/2+i\omega)t)$.
This result is well known and corresponds to the off-diagonal
elements of the density matrix in the coherent state basis in the
presence of a bath where $\rho(t=0)=|\alpha\rangle\langle\beta|$.

For a second example we take $\omega(t)=\omega_{0}(1+A\sin\,\eta t)$,
$f_{1}=ie^{i\omega_{d}t}D/2$, $\bar{f}_{1}=f_{1}^{*}$ and
$f_{2}=\bar{f}_{2}=0$.
This corresponds to a variable frequency oscillator with a driving
force. One again proceeds as before, however, the ODE's are slightly
more complicated due to the forcing term. The non-vanishing $g$'s are
\begin{eqnarray} g_{1} & = &
\frac{D/2(\exp(i\omega_{p}t)-\exp (-\gamma_{-}t-\epsilon(\cos\eta t-1)
))}{\gamma/2-i\delta\omega+\frac{\epsilon}{\eta}\sin\eta t}\;\;,\\
g_{2} & = &
\frac{D/2(\exp(-i\omega_{p}t)-\exp(-\gamma_{+}t+\epsilon(\cos\eta -1)
t))}{\gamma/2+i\delta\omega-\frac{\epsilon}{\eta}\sin\eta t}\;\;,\\
g_{7} & = & (2\bar{n}+1)e^{-\gamma t/2}\sinh \frac{\gamma t}{2}\;\;,\\
g_{8} & = & 2\gamma t \;\;,\\
g_{9} & = & -2i\omega_{0}t+2\epsilon(\cos\eta t -1)\;\;,
\end{eqnarray}
where $\gamma_{\pm}=\gamma/2\pm i\omega_{0}$,
$\delta\omega=\omega_{0}-\omega_{d}$ and $\epsilon=i\omega_{0}A/\eta$.
The propagator on the extended Hilbert-space is now
$U(t)=e^{g_{1}W_{1}}\,e^{g_{2}W_{2}}\,e^{g_{7}W_{7}}\,e^{g_{8}W_{8}}
\,e^{g_{9}W_{9}}e^{g_{0}W_{0}}$.
We again find $\dot{g_{0}}=\gamma /2$. To compute the result analogous
to (\ref{prop-alpha}) we normal order the $W_{7},W_{8}$ and $W_{9}$
exponentials and insert a resolution of unity in the Super--Hilbert
space between $e^{g_{2}W_{2}}$
and $e^{xK_{+}}$. To complete the calculation we perform the coherent
state integral over the introduced partition of unity using
$\int\frac{d^{2}\alpha}{\pi}\;e^{-A\alpha^{*}\alpha+B\alpha+C\alpha^{*}}=
\frac{1}{A}e^{BC}$. The final result is
\begin{eqnarray}\langle&&\tilde{\beta}^{*},\tilde{\alpha}|U(t)|\alpha,
\beta^{*}\rangle= \nonumber\\
&&\frac{1}{\kappa}\langle\tilde{\alpha}|\tilde{\beta}\rangle\langle\beta
|\alpha\rangle
\exp\left[-\frac{1}{\kappa}
(g_{1}+\tilde{\alpha}-\beta\chi)^{*}(g_{2}+\tilde{\beta}-\alpha\chi)
\right]\;\;,
\label{prop-alpha2}\end{eqnarray}
where
\[\ln\,\chi= -\gamma t/2+g_{9}/2=-\gamma
t/2-i\omega_{0}(t-\frac{A}{\eta}(\cos\eta t-1))\].

In the latter model $u_{11}=\exp(\gamma_{-}t+\epsilon[\cos\,\eta t-1])$ and
$u_{33}=\exp(-\gamma_{+}t+\epsilon[ \cos\,\eta t-1])$. Since $|u_{11}|\neq 0$
and $u_{33}|\neq 0$ for all $\epsilon$ and finite $t$ the resulting
propagator in both models is global.
This method can thus treat complicated time dependencies. It can also
be applied to very particular
nonlinear systems \cite{CHATURVEDI_SRINIVASAN:1991}.

\section{Acknowledgments}
The author thanks Dr. J. McCarthy and Prof. W.
Unruh for helpful discussions.

\appendix
\section{BCH Identities}
In this Appendix we derive the necessary BCH
identities to disentangle the action of the propagator $U(t)$ in
(\ref{prop-ansatz}) on the elements $N_{i}$ of the ideal $\cal I$.

{}From \cite{DATTOLI_GALLARDO:1988} we have the expansion
\begin{equation} e^{xA}\, B \,
e^{-xA}=B+x\,[A,B]+\frac{x^2}{2!}\,[A,\,[A,B]]+\cdots
\;\;,\label{bch-exp}\end{equation}
where $x$ is a c--number and $A$ and $B$ are elements of a Lie group.
For the case where the commutator of $A$ and $B$ is a c--number
we are left with the first two terms in the expansion (\ref{bch-exp}).
For the simplest non--trivial case where the commutation of
$A$ with $B$ closes onto $B$ itself, ie., $[A,B]=mB$ one can
show
\begin{equation} e^{x\,{\rm ad}\,A}B=B\,e^{mx}\;\;.\end{equation}
For the more complicated example of
$\exp (-g_{14}\,{\rm ad}\,W_{14})\,W_{i}$ the commutation does not
close on the first
iteration. Here we use Wilkox's method of parameter differentiation
\cite{WILKOX:1967}.

Letting
\begin{equation} G(x)\equiv e^{xA}\, B e^{-xA}\;\;,\label{G-x}\end{equation}
with
\begin{equation}
[A,B]=m_{1}B+m_{2}C\;\;,\qquad[A,C]=m_{3}B+m_{4}C
\;\;,\label{commutation}\end{equation}
we can obtain
\begin{equation} G^{\prime}(x)\equiv\frac{dG}{dx}=[A,G(x)]\;\;.
\label{diff}\end{equation}
We now assume that $G(x)$ is of the form $G(x)=a(x)A+b(x)B+c(x)C$.
With the initial condition $G(x=0)=B$ we see that $a(x=0)=c(x=0)=0$
and $b(x=0)=1$. Substituting this ansatz for $G(x)$ into (\ref{diff}) we
obtain differential equations for $a(x)$, $b(x)$, and $c(x)$.
Immediately we get $a(x)=0$ and the coupled set
\begin{equation} \left(\begin{array}{c} b \\ c \end{array}\right)^{\prime}=
\left[ \begin{array}{cc} m_{1} & m_{3} \\ m_{2} & m_{4} \end{array}\right]
\left(\begin{array}{c} b \\ c \end{array}\right)\;\;.\label{diff2}
\end{equation}
For the particular example $\exp (-g_{14}\,{\rm ad}W_{14})W_{1}$
the values for the $m_{i}$
can be read off Table \ref{table4} and yield  $m_{1}=-1/2$,
$m_{2}=-1$, $m_{3}=2$, $m_{4}=1/2$. Inserting these values into
(\ref{diff2}) and solving with the above initial conditions, gives
\begin{equation} e^{-g_{14}{\rm ad}\,W_{14}}\,W_{1}=
[\cos  \frac{g_{14}}{\chi} +\frac{\chi}{2}\sin \frac{g_{14}}{\chi}]\,W_{1} +
[\chi\sin\,\frac{g_{14}}{\chi}]\,W_{3}\;\;,\end{equation}
where $\chi=2/\sqrt{7}$. The results of the action of $W_{14}$  on the
 elements of the ideal are given in  Table \ref{table6}.

\widetext
\begin{table}
\squeezetable
\caption{We give the realisation of the two-photon subalgebra of
Sp(6,r) in terms of creation and annihilation operators $a^{\dagger},
\tilde{a}^{\dagger},a,\tilde{a}$.}
\begin{tabular}{ccccccccccccccc}
$ N_{0} $&$ N_{1} $&$ N_{2} $&$ N_{3} $&$ N_{4} $&$ N_{5} $&$ N_{6} $&
$ N_{7} $&$ N_{8} $&$ N_{9} $&$ N_{10} $&$ N_{11} $&$ N_{12} $&
$ N_{13} $&$ N_{14}$\\\hline
$ 1 $&$ a $&$ \tilde{a} $&$ a^{\dagger} $&$ \tilde{a}^{\dagger} $&
$ a\tilde{a} $&$ a^{2} $&$ \tilde{a}^{2} $&$ a^{\dagger}\tilde{a} $&
$ a\tilde{a}^{\dagger} $&$a^{\dagger\,2} $&$ \tilde{a}^{\dagger\, 2} $&
$ a^{\dagger}\tilde{a}^{\dagger} $&$ a^{\dagger}a+1/2 $&
$ \tilde{a}^{\dagger}\tilde{a}+1/2$\\
\end{tabular}
\label{table1}\end{table}

\widetext
\begin{table}
\squeezetable
\caption{Table of coefficients $h_{i}$, appearing in the `Super-Hamiltonian'
(\protect\ref{schro4}).}
\begin{tabular}{ccccccccccccccc}
$ h_{0} $&$ h_{1} $&$ h_{2} $&$ h_{3} $&$ h_{4} $&$ h_{5} $&$ h_{6} $&
$ h_{7} $&$ h_{8} $&$ h_{9} $&$ h_{10} $&$ h_{11} $&$ h_{12} $&$ h_{13} $&
$ h_{14}$\\\hline
$\gamma/2 $&$ f_{1} $&$ -f_{1}^{*} $&$ f_{1}^{*} $&$ -f_{1} $&
$ i\gamma(\bar{n}+1) $&$ f_{2} $&$ -f_{2}^{*} $&$ 0 $&$ 0 $&$ f_{2}^{*} $&
$ -f_{2}  $&$ i\gamma\bar{n} $&$ \omega-i\gamma\bar{n}-i\gamma/2 $&
$-\omega-i\gamma\bar{n}-i\gamma/2$\\
\end{tabular}
\label{table2}
\end{table}

\mediumtext
\begin{table}
\caption{We give the relation between the new generators $W_{i}$ and
the old generators $N_{i}$. Also given are the new coefficient functions
appearing in (\protect\ref{schro4}) and the corresponding two-photon operator.}
\begin{tabular}{cccc}
${\bf W_{i}} $&$ {\bf N_{i} } $&$ {\bf \tilde{h_{i}}} $& {\bf Operator }
\\\hline
$W_{0} $&$  N_{0} $&$ \gamma/2 $&$ 1 $\\
$W_{1} $&$ N_{1}-N_{4}  $&$ f_{1}$&$ a -\tilde{a}^{\dagger} $\\
$W_{2} $&$ N_{2}-N_{3}  $&$ -f_{1}^{*}$&$ \tilde{a} - a^{\dagger} $\\
$W_{3} $&$ N_{1}+N_{4}  $&$ 0 $&$ a + \tilde{a}^{\dagger} $\\
$W_{4} $&$ N_{2}+N_{3}  $&$ 0 $&$ \tilde{a} + a^{\dagger} $\\
$W_{5} $&$ N_{6}-N_{11} $&$ f_{2} $&$ a^{2} -\tilde{a}^{\dagger\, 2} $\\
$W_{6} $&$ N_{7}-N_{10} $&$ -f_{2}^{*}$&$ \tilde{a}^{\dagger\, 2}
-a^{\dagger\,2} $\\
$W_{7} $&$ N_{5}+N_{12}-(N_{13}+N_{14}) $&$ i\gamma(2\bar{n}+1)/2 $&
$a\tilde{a} +a^{\dagger}\tilde{a}^{\dagger}-(a^{\dagger}a+
\tilde{a}^{\dagger}\tilde{a}+1) $\\
$W_{8} $&$ (N_{5}-N_{12})/4 $&$ 2i\gamma $&$ (a\tilde{a}-
a^{\dagger}\tilde{a}^{\dagger})/4 $\\
$W_{9} $&$ (N_{13}-N_{14})/2 $&$ 2\omega $&$ (a^{\dagger}a-
\tilde{a}^{\dagger}\tilde{a})/2 $\\
$W_{10} $&$ N_{8} $&$ 0 $&$ a^{\dagger}\tilde{a} $\\
$W_{11} $&$ N_{9} $&$ 0 $&$ \tilde{a}^{\dagger}a $\\
$W_{12} $&$ N_{7}+N_{10} $&$ 0 $&$ \tilde{a}^{2} + a^{\dagger\,2} $\\
$W_{13} $&$ N_{6}+N_{11} $&$ 0 $&$ a^{2} + \tilde{a}^{\dagger\, 2} $\\
$W_{14} $&$ 2N_{5}+N_{12}-(N_{13}+N_{14})/2 $&$ 0 $&$ 2a\tilde{a}
+a^{\dagger}\tilde{a}^{\dagger}-(a^{\dagger}a
+\tilde{a}^{\dagger}\tilde{a}+1)/2 $\\
\end{tabular}
\label{table3}\end{table}

\widetext
\begin{table}
\squeezetable
\caption{ The commutation table for the ideal
${\cal I}=\{W_{0},W_{1},W_{2},W_{3},W_{4}\}$ with the whole two-photon
group $\cal L$. The entries are $[W_{i},W_{j}]$.}
\begin{tabular}{c|cccccccccccccc}
$W_{i} \backslash W_{j} $&$ W_{1} $&$ W_{2} $&$ W_{3} $&$ W_{4} $&
$ W_{5 } $&$ W_{6} $&$ W_{7} $&$W_{8} $&$ W_{9} $&$ W_{10} $&$ W_{11} $&
$ W_{12} $&$ W_{13} $&$ W_{14} $\\\hline
$W_{1} $&$ 0 $&$ 0 $&$ 0 $&$ 2W_{0} $&$ 0 $&$ 2W_{2} $&$ 0 $&
$W_{1}/4 $&$W_{1}/2$&$ W_{4} $&$ 0 $&$ 2W_{4} $&$ 0 $&$ W_{1}/2+W_{3}  $\\
$ W_{2} $&$ 0  $&$ 0 $&$ 2W_{0} $&$ 0 $&$ 2W_{1} $&$ 0 $&$ 0 $&
$ W_{2}/4 $&$-W_{2}/2 $&$ 0 $&$ W_{3} $&$ 0 $&$ 2W_{3} $&$ W_{2}/2+W_{4}  $\\
$W_{3} $&$ 0 $&$ -2W_{0} $&$ 0 $&$ 0 $&$ 0 $&$ -2W_{4} $&$ -2W_{1} $&
$-W_{3}/4 $&$ W_{3}/2 $&$ W_{2} $&$ 0 $&$ -2W_{2} $&$  0 $&$-2W_{1}-W_{3}/2 $\\
$W_{4} $&$ -2W_{0} $&$ 0 $&$ 0 $&$ 0 $&$ -2W_{3} $&$ 0 $&$ -2W_{2} $&
$-W_{4}/4 $&$ -W_{4}/2 $&$ 0 $&$ W_{1} $&$ 0 $&$ -2W_{1} $&$-2W_{2}-W_{4}/2
$\\
\end{tabular}
\label{table4}\end{table}

\mediumtext
\begin{table}
\caption{Table of the action of the group element $W_{i}$ on an
element $W_{j}$ of the ideal $\cal I$, i.e. each entry shows
$\exp(-g_{i}W_{i})W_{j}\exp(g_{i}W_{i})$.}
\begin{tabular}{l|llll}
$W_{i}\backslash W_{j} $&$ W_{1} $&$ W_{2} $&$ W_{3} $&$ W_{4} $\\ \hline
$W_{1} $&$ W_{1}$&$ W_{2}$&$ W_{3}$&$ W_{4}-2g_{1} $\\
$W_{2} $&$ W_{1}$&$ W_{2}$&$ W_{3}-2g_{2}$&$ W_{4} $\\
$W_{3} $&$ W_{1}$&$ W_{2}+2g_{3}$&$ W_{3}$&$ W_{4} $\\
$W_{4} $&$ W_{1}+2g_{4}$&$ W_{2}$&$ W_{3}$&$ W_{4} $\\
$W_{5} $&$ W_{1}$&$ W_{2}+2g_{5}W_{1} $&$ W_{3} $&$ W_{4}-2g_{5}W_{3} $\\
$W_{6} $&$ W_{1}+2g_{6}W_{2} $&$ W_{2} $&$ W_{3}-2g_{6}W_{4} $&$ W_{4} $\\
$W_{7} $&$ W_{1} $&$ W_{2} $&$ W_{3}-2g_{7}W_{1} $&$ W_{4}-2g_{7}W_{2} $\\
$W_{8} $&$ W_{1}\,e^{g_{8}/4}  $&$ W_{2}\,e^{g_{8}/4}$&$ W_{3}
\,e^{-g_{8}/4} $&$ W_{4}\,e^{-g_{8}/4} $\\
$W_{9} $&$ W_{1}\,e^{g_{9}/2}  $&$ W_{2}\,e^{-g_{9}/2}$&$ W_{3}
\,e^{g_{9}/2}  $&$ W_{4}\,e^{-g_{9}/2} $\\
$W_{10}$&$ W_{1}+g_{10}W_{4}   $&$ W_{2}$&$ W_{3}+g_{10}W_{2}   $&
$ W_{4} $\\
$W_{11}$&$ W_{1}     $&$ W_{2}+g_{11}W_{3} $&$ W_{3}    $&$ W_{4}
+g_{11}W_{1} $\\
$W_{12}$&$ W_{1}+2g_{12}W_{4}$&$W_{2}$&$W_{3}-2g_{12}W_{2}  $&$ W_{4} $\\
$W_{13}$&$ W_{1} $&$W_{2}+2g_{13}W_{3}$&$ W_{3}$&$W_{4}-2g_{13}W_{1}$\\
\end{tabular}
\label{table5}\end{table}

\narrowtext
\begin{table}
\caption{Table of the action of $W_{14}$ on the elements of the ideal
$\cal I$.
We have set $\tan\, \mu_{1}=7^{-1/2}$ and $\mu_{2}=g_{14}/\alpha$
where $\alpha=2/7^{1/2}$.
See the Appendix for more details.}
\begin{tabular}{c|c}
 & $W_{14}$\\ \hline
$W_{1}$&$\alpha(\sqrt{2}\cos(\mu_{1}-\mu_{2})W_{1}+\sin\mu_{2}W_{3}) $\\
$W_{2}$&$\alpha(\sqrt{2}\cos(\mu_{1}-\mu_{2})W_{2}+\sin\mu_{2}W_{4}) $\\
$W_{3}$&$\alpha(\sqrt{2}\cos(\mu_{1}+\mu_{2})W_{3}-2\sin\mu_{2}W_{1}) $\\
$W_{4}$&$\alpha(\sqrt{2}\cos(\mu_{1}+\mu_{2})W_{4}-2\sin\mu_{2}W_{2})$\\
\end{tabular}
\label{table6}\end{table}

\mediumtext
\begin{table}
\caption{Algebraic relations between the coefficients $g_{i}$ and the
functions $u_{ij}$.}
\begin{tabular}{lll}
$g_{1}=-u_{40}/2 	$&$ g_{2}=-u_{30}/2 	$&$ g_{3}=0 $\\
$g_{4}=0 		$&$ g_{5}=u_{21}/(2u_{11})$&$ g_{6}=u_{12}
/(2u_{11})$ \\
$g_{7}=-u_{31}/(2u_{11})$&$ g_{8}=2\ln(u_{11}/u_{33}) $&$ g_{9}=
\ln(u_{11}u_{33}) $\\
$g_{10}=-2g_{12} $&$ g_{11}=-2g_{13} $&$ g_{12}=-\left[ u_{32}
-u_{31}u_{12}/u_{11}\right]/(4u_{33})$ \\
$g_{13}=-\left[u_{41}+u_{21}u_{31}/u_{11}\right]/(4u_{11}) $& &  \\
\end{tabular}
\label{table7}\end{table}

\end{document}